\documentclass[aps,prd,notitlepage,nofootinbib,preprintnumbers,superscriptaddress]{revtex4}
\usepackage{amssymb}
\usepackage{amsmath}
\usepackage{verbatim}
\usepackage{appendix}
\usepackage[usenames,dvipsnames]{xcolor}

\begin{document}

\title{Stochastic approach to de Sitter instability and eternal inflation}

\author{Rio Saitou}
\email{riosaitou18@hotmail.com}
\affiliation{Leung Center for Cosmology and Particle Astrophysics, National Taiwan University, Taipei, Taiwan 10617}

\begin{abstract}

We investigate when effective theories of a scalar field on (quasi-)de Sitter background break down through the stochastic formalism. 
We derive the Fokker-Planck equation leaving the second order time derivative of the scalar field.
Assuming there exists an equilibrium distribution for the field velocity, we obtain a mean value and a variance of the field velocity caused by the quantum fluctuation. Introducing coarse-grained Einstein equations, we obtain bounds for the non-eternal inflation phase and for maintaining the exact de Sitter background.
We point out that those bounds derived in our formalism correspond to the de Sitter entropy bound proposed by Arkani-Hamed, \textit{et.al.}, up to $O(1)$ factor, even for a massless free scalar field on exact de Sitter background.
We discuss connections of our results to the quantum field theory also.

\end{abstract}

\maketitle

\section{Introduction}

We describe the evolution of the universe, especially the inflationary universe, by the cosmological perturbation theory or scalar fields on de Sitter background (See, for example, \cite{Kodama:1985bj, Mukhanov:1990me, Copeland:2006wr} and \cite{Birrell} also). Those theories are defined on  classical backgrounds characterized by the Hubble parameter $H$. It has been suggested that those effective theories on the cosmological backgrounds will break down by quantum effects \cite{Baumann:2011ws, Mottola:1984ar, Mottola:1985qt, Tsamis:1992sx, Goheer:2002vf, Polyakov:2007mm, Anderson:2013zia, Markkanen:2016jhg, Dvali:2017eba, Markkanen:2017abw, Matsui:2019tah, Aalsma:2019rpt}.
In \cite{Cheung:2007st, Baumann:2011ws}, it was clearly shown that the conserved curvature perturbation in single-field inflation
models is the Nambu-Goldstone (NG) boson associated with the time translation symmetry breaking. The authors in \cite{Baumann:2011ws} indicate that if the size of NG boson exceeds the cutoff scale of the NG phase, the effective field theory breaks down and the system enters in the eternal inflation phase \cite{Vilenkin:1983xq, Linde:1986fd}. 
On the other hand, the scalar field theory on de Sitter background, which is not in the NG phase,  get constrained also. The approaches based on the quantum field theory \cite{Mottola:1984ar, Mottola:1985qt, Tsamis:1992sx, Goheer:2002vf, Polyakov:2007mm, Anderson:2013zia, Markkanen:2016jhg, Dvali:2017eba, Markkanen:2017abw, Matsui:2019tah, Aalsma:2019rpt} suggested that de Sitter spacetime will collapse even in a far lower energy scale than the Planck scale. 

%
To include quantum effects in cosmology, a simple but useful method has been established as the stochastic formalism \cite{Starobinsky:1986fx, Goncharov:1987ir, Salopek:1990re, Starobinsky:1994bd, Tolley:2008na, Rigopoulos:2013exa, Garbrecht:2013coa, Garbrecht:2014dca, Vennin:2015hra, Rigopoulos:2016oko, Moss:2016uix, Prokopec:2017vxx, Pinol:2018euk}.
In the formalism, we divide a scalar field to a long and a short wavelength parts. 
The long wavelength part which is responsible for the cosmological background gets corrected by the short wavelength part superposed on the long part. We usually approximate the short wavelength part as white noise, and the long wavelength part starts to have a variance as a quantum effect from the noise. This would correspond to a coarse-graining or a renormalization for low frequency modes in the quantum field theory. Indeed, several approaches based on the quantum field theory reproduce the same results as the stochastic formalism in a proper limit \cite{Garbrecht:2013coa, Garbrecht:2014dca}. If the approximations in the stochastic formalism are appropriate, however, we can elicit intrinsic results from the stochastic formalism only. Thus, we expect that the stochastic formalism is sufficient to see the quantum effects, as the first order approximation of the full quantum field theory. 

In this article, we apply the stochastic formalism to a system consisting of a canonical scalar field
$\phi$ and the general relativity with clarifying approximations. We investigate a valid range for the effective theory on the classical (quasi-)de Sitter background.
Intuitively, we can expect to see deviations from purely classical results since we take into account the coarse-graining effect of the short wavelength part to the classical background.
Unlike the usual way, we do not use the slow roll approximation for the Klein-Gordon equation and derive the Fokker-Planck equation leaving the second order time derivative of the scalar field.
Then, we derive an equilibrium solution for the velocity distribution, associated mean value, and variance of the field velocity. Moreover, we simply perform the coarse-graining for the Einstein equations also. Using them, we investigate how the classical background get corrected by the quantum effect of the short wavelength part. We consider two different cases. One is the slow roll inflation in which we give the classical background as quasi-de Sitter spacetime. Another one is the massless free scalar field on exact de Sitter background made of a constant potential of $\phi$. From our approach based on the coarse-grained Einstein equations, we will (re)find that those effective field theories on the classical backgrounds are valid only when the following bound is satisfied:
\begin{align}
\label{bound}
    &S_{\rm dS}:=\frac{\pi}{GH^2} \gtrsim N \ , 
\end{align}
where $G$ is the gravitational constant and $N$ is the e-folding number. This bound just corresponds to the de Sitter entropy bound conjectured in \cite{ArkaniHamed:2007ky}.

The layout of this article is as follows: In Sec. \ref{2}, we review the stochastic formalism in detail.
We give the equilibrium solution for the velocity distribution and consider the thermodynamical meaning of the distribution. In Sec. \ref{4}, we investigate the evolution of the system using the coarse-grained Einstein equations for the two different cases. First, we consider
the case where the system undergoes the slow roll inflation. 
We will find that out of the bound (\ref{bound}), the system transit to the eternal inflation phase. Then, we consider the massless free scalar field with the constant potential where the classical background is given by exact de Sitter spacetime. We will (re)find that the de Sitter spacetime evaporates and the effective field theory breaks down when the bound (\ref{bound}) is no longer kept.  In Sec. \ref{5}, we give concluding remarks and discuss the connection of our results to the quantum field theory. We calculate a correlation of the short wavelength part in Appendix. 

\section{Stochastic formalism}\label{2}

\subsection{Stochastic inflation}
We introduce the stochastic formalism step by step.
We consider a single canonical scalar field $\phi$ with the general relativity. The action is given by
\begin{align}
\label{}
    S= \int d^4x\sqrt{-g}\left[ \frac{M_{\rm pl}^2}{2}R - \frac{1}{2}g^{\mu\nu}\nabla_\mu\phi\nabla_\nu\phi - V(\phi)\right]   \ ,
\end{align}  
where $R$ is a scalar curvature, $\nabla_{\mu}$ is a covariant derivative associated with a metric $g_{\mu\nu}$ and $V(\phi)$ is a potential of $\phi$. Variating with respect to $\phi$, we obtain the Klein-Gordon (KG) equation
\begin{align}
\label{}
    &\Box\phi + \partial_\phi V(\phi)= 0,   \quad \Box:=-g^{\mu\nu}\nabla_\mu\nabla_\nu \ ,
\end{align} 
where $\partial_\phi:= \partial/\partial\phi$. Throughout this article, $\partial_\Psi$ will denote a partial derivative of a variable $\Psi$. 
We impose the flat Friedmann-Lema\^{\i}tre-Robertson-Walker (FLRW) metric for the gravity sector
\begin{align}
\label{}
    ds^2&= -dt^2 + a^2(t)d\vec{x}^2 \ .
\end{align} 
We introduce e-folding number $N$ as the time variable instead of $t$
\begin{equation}
\label{ }
  dN = Hdt,\quad N = {\rm ln}\left(\frac{a}{a_0}\right) \ ,
\end{equation}
where $H:= d{\rm ln}a/dt$ is the Hubble parameter and $a_0$ is a value of scale factor at $N=0$. 
We canonically quantize the scalar field on the flat FLRW metric
\begin{align}
\label{}
    &\pi_\phi:= \frac{\delta S}{\delta( \partial_0 \phi)} = -\sqrt{-g}g^{0\nu}\nabla_\nu \phi =
           a^3H\partial_N\phi  \ , \nonumber \\
    &[\hat{\phi}(N, \mathbf{x}),\,\hat{\pi}_\phi(N,\, \mathbf{y})]= i\delta^3(\mathbf{x}-\mathbf{y})   \ ,
\end{align}
where the variables with hats denote the Heisenberg operators.  

Since we can write the operator as a superposition of waves, we can split $\hat{\phi}$ linearly to long and short wavelength parts as
\begin{align}
\label{divide}
    &\hat{\phi}(N,\mathbf{x}) = \hat{\phi}_L(N,\mathbf{x}) + \hat{\phi}_S(N,\mathbf{x})  \ . 
\end{align}
The subscripts $L$ and $S$ implies the long and short wavelength parts respectively. 
This procedure is compatible with decomposing the wave function to in and out sates. We impose the following approximations for the system;
\begin{enumerate}
  \item We ignore all the components of metric perturbation.
  \item The Heisenberg operator $\hat{\phi}$ obeys to the Klein-Gordon equation on the flat FLRW metric.
\end{enumerate}
The first one is reasonable for the scalar components of metric perturbation since they are small compared with the scalar field perturbation in sub horizon scales \cite{Cheung:2007st, Baumann:2011ws}. We ignore, however, the tensor mode, i.e. graviton also, while graviton does not affect the evolution of the scalar perturbation at the linear order. Hence, in this article, we allow ignoring the effect of graviton. 
%
Assuming the full operator obeys to the KG equation, we obtain
\begin{align}
\label{KGL}
    \Box\hat{\phi}_L + \partial_{\phi} V(\hat{\phi}_L) &= -\left[\Box +\partial_{\phi}^{2}V(\hat{\phi}_L)\right]\hat{\phi}_S 
        - \sum_{n=2}^\infty \frac{1}{n!}\partial_{\phi}^{n+1}V(\hat{\phi}_L)\,\hat{\phi}_S^n \ .
\end{align}
$\partial_\phi^n$ denotes the $n$-th order derivative of $\phi$.
According to the $\delta N$ formalism \cite{Lyth:2004gb}, the long wavelength part will create the perturbed FLRW spacetime\footnote{
With a classicalization of the long wavelength part.}. 
Each local Hubble patch, however, is still regarded as the flat FLRW spacetime. 
We focus on one of such local Hubble patches and consider the effect of short wavelength part superposed on our patch. The classical background equations generated by the long wavelength part are given as
%
%
\begin{align}
\label{Feq}
    3M_{\rm pl}^2H^2 &\overset{\rm cl.}{=} \frac{1}{2}H^2(\partial_N\bar{\phi})^2 + V(\bar{\phi}) \ ,  \\
    2M_{\rm pl}^2H\partial_NH &\overset{{\rm cl.}}{=} -H^2(\partial_N\bar{\phi})^2 \ ,
\end{align}
where `${\rm cl.}$' denotes the equations of motion are classical, and $\bar{\phi}$ is a classical field configuration on the local Hubble patch.
For the slow roll potential, we approximately obtain
\begin{align}
\label{esti}
    &3M_{\rm pl}^2H^2\overset{\rm cl.}{\simeq} V(\bar{\phi}) \ , \nonumber \\
    &\partial_N \bar{\phi}\overset{\rm cl.}{\simeq} -\frac{\partial_\phi V(\bar{\phi})}{3H^2} \ .
\end{align}
%
We define slow roll parameters as 
\begin{align}
\label{}
    &\epsilon_{1}:= -\partial_N{\rm ln}H \ ,\nonumber    \\
    &\epsilon_{i+1}:= \partial_N{\rm ln}\epsilon_i,\quad i\geq 1 \ .
\end{align}
We assume that all of the slow roll parameters are smaller than $1$:  $|\epsilon_i| < 1$, that is, we consider the (quasi-)de Sitter spacetime only.
The condition $|\epsilon_i| < 1$ corresponds to the adiabatic condition for the gauge-invariant curvature perturbation under which it always conserves on super horizon scales. 
Then, using Eqs. (\ref{esti}), we can estimate the size of derivatives of potential in Eq. (\ref{KGL}) as
\begin{align}
\label{}
    \partial_{\phi}^{\,n}V(\hat{\phi}_L) \sim M_{\rm pl}^{2-n}H^2\cdot O\left(\epsilon_i^{n/2}\right) \ . 
\end{align}
%
%
Therefore, for the generic potential satisfying the adiabatic condition, we can roughly estimate that all of the interactions are sufficiently weak to ignore. Thus, as the third approximation, we impose
\begin{enumerate}
  \setcounter{enumi}{2}
  \item We ignore all of the interactions of the short wavelength part and thus we treat it as a free scalar field on the (quasi-)de Sitter spacetime.
\end{enumerate}
This can be interpreted as the first order approximation for the interaction theory on the (quasi-)de Sitter spacetime.

According to the third approximation, we perform the mode expansion accompanied by the creation and annihilation operators $\hat{a}_{\mathbf{k}}^{(\dag)}$ for the short wavelength part only 
\begin{align}
\label{split}
    \hat{\phi}(N,\mathbf{x}) &= \hat{\phi}_L(N,\mathbf{x}) + \hat{\phi}_S(N,\mathbf{x}) \nonumber \\
    &=\hat{\phi}_L(N,\mathbf{x}) + \int \frac{d^3k}{(2\pi)^{3/2}} \theta(k-\varepsilon aH)
         \,\hat{\varphi}_\mathbf{k}{\rm e}^{i\mathbf{k}\cdot\mathbf{x}} \ ,  \\ 
    \hat{\varphi}_\mathbf{k}&:= \hat{a}_{\mathbf{k}}\varphi_k+ \hat{a}_{-\mathbf{k}}^\dag \varphi_k^* \ , \nonumber
\end{align}
where $\theta$ is the Heaviside step function satisfied with
\begin{align}
\label{}
    \theta(z)= \begin{cases}
      1& (z> 0), \\
      0& (z< 0).
\end{cases} \nonumber 
\end{align}
The vacuum is satisfied with 
\begin{equation}
\label{ }
   \hat{a}_{\mathbf{k}}|0\rangle = 0 \ .
\end{equation}
The small parameter $\varepsilon$ specifies that the short wavelength part includes almost the sub horizon modes only, and this will not appear in the final result.
The short wavelength part represents almost the sub horizon modes of linear cosmological perturbation,  
so that the mode function $\varphi_k$ is satisfied with the usual perturbed field equation for the cosmological perturbation
\begin{align}
\label{MS}
    &[\Box +\partial_{\phi}^2V(\bar{\phi})]\varphi_k{\rm e}^{i\mathbf{k}\cdot\mathbf{x}} \nonumber \\
   =&\left[\partial^2_t\varphi_k + 3H\partial_t \varphi_k +\frac{k^2}{a^2}\varphi_k + \partial_{\phi}^2V(\bar{\phi})\varphi_k\right]{\rm e}^{i\mathbf{k}\cdot\mathbf{x}} \nonumber \\
   =& 0  \ .
\end{align}
We note that the short wavelength part itself is \textit{not} a usual cosmological perturbation since it includes the step function as a separation between in and out states, that is, the short wavelength part has a smaller support than the usual cosmological perturbation in the momentum space.
%
Discarding the interaction terms of short wavelength part,
we can reduce the full Klein-Gordon equation (\ref{KGL}) to
\begin{align}
\label{KGL2}
    \Box\hat{\phi}_L + \partial_{\phi}V(\hat{\phi}_L) &\simeq -[\Box + \partial_{\phi}^2V(\bar{\phi})]\hat{\phi}_S \ .
\end{align}
Up to the leading order of the slow roll parameters, the right hand side of Eq. (\ref{KGL2}) becomes 
\begin{align}
\label{noi}
    -[\Box + \partial_{\phi}^2V(\bar{\phi})]\hat{\phi}_S &\simeq -\int \frac{d^3k}{(2\pi)^{3/2}}
        \left[H^2(\partial_N^2 \theta + 3\partial_N \theta + 2\partial_N\theta \partial_N)
          + \theta \cdot (\Box + \partial_{\phi}^2V(\bar{\phi}))\right]
          \hat{\varphi}_\mathbf{k}{\rm e}^{i\mathbf{k}\cdot \mathbf{x}} \nonumber \\
          &\simeq  H^2\int \frac{d^3k}{(2\pi)^{3/2}}
           \left\{\partial_N\left[ \varepsilon aH\cdot \delta (k-\varepsilon aH)
             \hat{\varphi}_\mathbf{k}\right] \right. \nonumber \\
          &\qquad \quad+ \left. \left[\varepsilon aH \cdot \delta (k-\varepsilon aH)\right]
             (3\hat{\varphi}_\mathbf{k} +\partial_N\hat{\varphi}_\mathbf{k})\right\}
                {\rm e}^{i\mathbf{k}\cdot \mathbf{x}}  \ ,
\end{align}
where we use Eq. (\ref{MS}) for the second equality.
If we choose the Bunch-Davies vacuum, the leading solution of Eq. (\ref{MS}) is given as
\begin{align}
\label{}
    \varphi_k&\simeq \frac{iH}{\sqrt{2k^3}}(1+ik\tau){\rm e}^{-ik\tau}  \ , \nonumber  \\
    \tau &= -\frac{1}{aH}\ ,
\end{align}
which  behaves almost as a constant on the super horizon scales
\begin{align}
\label{}
    &\varphi_{k=\varepsilon aH}\simeq \frac{iH}{\sqrt{2k^3}} \simeq {\rm constant} \ ,  \\
    &\partial_N\varphi_{k=\varepsilon aH} \simeq 0 \ .
\end{align}
Then, the right hand side of Eq. (\ref{KGL2}) finally becomes 
\begin{align}
\label{}
    -[\Box + \partial_{\phi}^2V(\bar{\phi})]\hat{\phi}_S&\simeq H^2(\partial_N + 3)\int \frac{d^3k}{(2\pi)^{3/2}}\varepsilon aH\cdot \delta(k-\varepsilon aH)\frac{iH}{\sqrt{2k^3}}[\hat{a}_\mathbf{k}{\rm e}^{i\mathbf{k}\cdot\mathbf{x}} - \hat{a}_\mathbf{k}^\dag {\rm e}^{-i\mathbf{k}\cdot\mathbf{x}}]  \nonumber \\
    &=: H^2\hat{\xi}_S(N,\mathbf{x})   \ .
\end{align}
This term appears since the support of operator $\hat{\phi}$ is divided by the step function $\theta$.


We can safely ignore the spatial gradient term of the long wavelength part from the KG equation (\ref{KGL2}) since the long wavelength part includes highly super horizon modes only.
Then, up to the leading order, we obtain 
\begin{align}
\label{Lan}
    &\hat{v}_L(N,\mathbf{x})= \partial_N\hat{\phi}_{L}(N,\mathbf{x}) \ ,   \nonumber \\
    &\partial_N\hat{v}_{L} = - 3\hat{v}_L - \frac{\partial_{\phi}V(\hat{\phi}_L)}{H(N)^2} + \hat{\xi}_S(N,\mathbf{x}) \ .
\end{align}
Unlike the classical KG equation for $\bar{\phi}$, we receive a contribution from the short wavelength part. 
We write $V(\hat{\phi}_L)$ as $\hat{V}$ and omit ${\bf x}$ from the arguments hereafter.
Taking differences, we get
\begin{align}
\label{}
    \int^{N+\Delta N}_N \partial_{\tilde{N}}\hat{v}_{L}d\tilde{N}&= \hat{v}_L(N+\Delta N) -\hat{v}_L(N) 
     =: \Delta \hat{v}_L   \\
    \int^{N+\Delta N}_N \left( -3\hat{v}_L -\frac{\partial_{\phi}\hat{V}}{H^2} + \xi_S\right)d\tilde{N}&= \left( -3\hat{v}_L -\frac{\partial_{\phi}\hat{V}}{H^2}\right )\Delta N + \Delta \hat{W}\ , \\
    \Delta \hat{W} &:= \int^{N+\Delta N}_N \hat{\xi}_S(\tilde{N}) d\tilde{N}  
\end{align}
We can derive moments of $\Delta W$ at the same points by taking vacuum expectation values as 
\begin{align}
\label{white}
    &\langle \Delta \hat{W}\rangle = 0   \ , \nonumber \\
    &\langle \Delta \hat{W}^2\rangle \simeq \left(\frac{3H}{2\pi}\right)^2\Delta N   \ .
\end{align}
See Appendix A for the technical detail of derivation. 
We ignore the interactions for the short wavelength part so that $\Delta \hat{W}$ must have a Gaussian distribution. Then, if we regard the Hubble parameter as a constant, we can identify Eq. (\ref{white}) to white noise and hence regard Eq. (\ref{Lan}) just as a Langevin equation. The reason why we receive the white noise is that the short wavelength part freezes out after leaving the horizon. For the Minkowski spacetime, we would receive a Gaussian noise, but not white. Thus, the (approximate) white noise arises from the introduction of (quasi-)de Sitter spacetime with the condition $|\epsilon_i|< 1$. 

Following the usual procedure, we can derive the Fokker-Planck equation as
\begin{align}
\label{PDF}
    &\partial_N P(v_L, \phi_L, N) = -\partial_{\phi_L} J_{\phi_L} - \partial_{v_L} J_{v_L} \ ,  \\
    &J_{\phi_L} :=  v_LP\ ,\nonumber \\
    &J_{v_L} := \left(-3v_L-\frac{\partial_{\phi}V(\phi_L)}{H^2} - \frac{D}{2}\partial_{v_L}\right)P \ ,\nonumber \\
    &D:= \frac{9H^2}{4\pi^2} \ .\nonumber 
\end{align}
We imposed that the probability distribution function (PDF) $P$ and its time derivative fall off at the boundary. The diffusion coefficient $D$ differs from the usual stochastic formalism since we consider the velocity distribution also. We will find in the next section that if we have an equilibrium state for the velocity distribution, we recover the usual value of diffusion coefficient.
%
%

\subsection{(quasi-)de Sitter equilibrium}\label{3}

We assume that the velocity distribution approaches to an equilibrium state.
This assumption is reasonable for the slow roll inflation model since it has the slow roll attractor at the classical level where we can ignore the time derivative of field velocity. For the ultra slow roll inflation model, at the classical level, the time derivative of field velocity is proportional to the field velocity which quickly approaches to $0$: $\partial_t\bar{\phi}\to 0$. Then, we can still imagine there exists an equilibrium state for the velocity distribution.

We can obtain the equilibrium state of velocity distribution supposing the zero velocity flux 
\begin{align}
\label{}
    &J_{v_L,\rm eq} = \left(-3v_L-\frac{\partial_{\phi}V}{H^2} - \frac{D}{2}\partial_{v_L}\right)P_{v_L,\rm eq} = 0  \ . 
\end{align}
Solving this with respect to $v_{L}$, we obtain the static velocity distribution as
\begin{align}
\label{Peq}
    &P_{v_L,\rm eq} \propto {\rm exp}\left[ -\frac{3}{D}\left( v_L+ \frac{\partial_{\phi}V}{H^2}\right)^2\right] \ .  
\end{align}
The mean value of field velocity becomes
\begin{align}
\label{VD}
    &\langle \hat{v}_L \rangle = \langle \partial_N\hat{\phi}_L\rangle = -\frac{\partial_{\phi}V(\phi_L)}{H^2}  \ , 
\end{align}
which is the drift of field velocity \cite{Prokopec:2017vxx}.
This corresponds to the classical attractor solution for the slow roll inflation as expected. If the potential is a constant, the mean velocity becomes zero and we have the exact de Sitter spacetime as the classical background.  
We note that in the right hand side of Eq. (\ref{VD}), the field is no longer a random variable but has a measured value.
The variance becomes
\begin{align}
\label{}
    \sigma_v^2 &= \frac{D}{6} = \frac{3H^2}{8\pi^2} \ , \\
    \label{sigma}
    \langle (\partial_N\hat{\phi}_{L})^2\rangle &= \langle \partial_N\hat{\phi}_{L}\rangle^2 + \sigma_v^2 = \left(\frac{\partial_{\phi}V(\phi_L)}{H^2}\right)^2 + \frac{3H^2}{8\pi^2} \ ,
\end{align}
We obtained the additional term besides the classical value for the kinetic term of field, which is generated from the quantum fluctuation of the short wavelength part. This effect would correspond to a coarse-graining or a renormalization for the long wavelength part in the thermal field theory with the Gibbons-Hawking temperature $T_H=H/2\pi$.

When the potential $V$ is the slow roll potential, we can substitute $\partial_N\hat{v}_L=0$ to the Langevin equation (\ref{Lan}) since the velocity is in the equilibrium state. Then, we obtain
\begin{align}
\label{}
    &\frac{d\hat{\phi}_L}{dN} =  -\frac{\partial_{\phi}V(\hat{\phi}_L)}{3H^2} + \frac{\hat{\xi}_S(N)}{3} \ ,
\end{align}
which corresponds to the usual Langevin equation for the stochastic inflation. From this, we can derive the Einstein-Smoluchowski equation where the diffusion coefficient is given by $\tilde{D}=\left(H/2\pi\right)^2$.

The velocity distribution of the long wavelength part becomes a classical distribution since the field classicalizes and loses its quantum properties. Hence, the velocity distribution in the equilibrium state will become the Boltzmann distribution. The kinetic energy of `particle' part of $\hat{\phi}_L$ is given by
\begin{align}
\label{}
    &K = \frac{1}{2}\left( \partial_t\phi_L-\langle \partial_t \hat{\phi}_L\rangle\right)^2\cdot L^3=  \frac{1}{2}H^2(v_L-\langle \hat{v}_L\rangle)^2\cdot L^3 \ ,
\end{align} 
where $L^3$ is a relevant spatial volume. The Boltzmann distribution requires
\begin{align}
\label{}
    &{\rm exp}[-\beta K] =  {\rm exp}\left[ -\frac{3}{D}\left( v_L -\langle \hat{v}_L\rangle\right)^2\right]  \ ,\\
    &\beta := \frac{1}{k_B T} \ .
\end{align}
If we take the Hubble volume $L_H^3 := 4\pi/3H^3$ and 
give the temperature as $T_H = H/2\pi$, we obtain the following relation
\begin{align}
\label{}
    &D =  \frac{2\cdot 3 k_BT_H}{H^2L_H^3} = \frac{9H^2}{4\pi^2} \ , 
\end{align}
for $k_B=1$.
This is the second fluctuation-dissipation theorem (2nd FDT) for the long wavelength part in the (quasi-)de Sitter spacetime \cite{Rigopoulos:2013exa, Rigopoulos:2016oko}. 

\section{Coarse-grained Einstein equation}\label{4}

The long wavelength part itself has a classical configuration $\bar{\phi}$ which generates the (local) FLRW spacetime at each Hubble patch via the classical Einstein equations (\ref{Feq}). Now, we include the short wavelength part superposed on the long wavelength part as the white noise. We should be able to evaluate the effect of the noise to the classical Einstein equations by taking an average for the long wavelength part of the energy-momentum tensor with the probability distribution function $P$.
The noise originates from the short modes of quantum fluctuation so that the effect from the noise represents a quantum correction from the short part to the long part.  
This procedure would corresponds to a coarse-graining or a renormalization to the low frequency modes in the quantum field theory.
We approximate the gravity sector by the flat FLRW metric since we assume we can ignore the metric perturbation.

Taking the average for the energy-momentum tensor, we introduce coarse-grained Einstein equations as follows:
\begin{align}
\label{}
    3M_{\rm pl}^2H^2& = \left\langle \frac{1}{2}H^2(\partial_N\hat{\phi}_{L})^2 + V(\hat{\phi}_L)\right\rangle \ ,  \\
    2M_{\rm pl}^2HH_{,N} &= -\left\langle H^2(\partial_N\hat{\phi}_{L})^2\right\rangle  \ .
\end{align}
We note that to get the average in the first equation, we need a full PDF including the distribution of the field value also if the potential depends on the field value. On the other hand, for the average in the second equation, it is sufficient to know the velocity distribution only. We evaluate the average over the velocity distribution for two different cases using the equilibrium distribution (\ref{Peq}). 


\subsection{Eternal inflation}

We consider the single-field slow roll inflation induced by a slow roll potential $V(\phi)$ in the stochastic formalism. We evaluate the second coarse-grained Einstein equation only with use of Eqs. (\ref{Peq}) and (\ref{sigma}) :
\begin{align}
\label{2nd}
    2M_{\rm pl}^2HH_{,N} 
       &=-\langle \partial_t{\hat{\phi}}_L\rangle^2
             - \frac{3H^4}{8\pi^2}    
\end{align}
The first term in the right hand side is the square of the mean value for the long wavelength part and thus this corresponds to the classical value at the local Hubble patch: $\langle \partial_t{\hat{\phi}}_L\rangle = \partial_t{\bar{\phi}}$. The second term is the variance of long wavelength part created from the quantum fluctuation (noise) of the short wavelength part. 
We calculated the evolution of short wavelength part on the classical background given by Eq. (\ref{Feq}). Thus, we have to require that the background spacetime is determined by the classical value of the scalar field, otherwise, the effective field theory will break down. This imposes the following condition to the theory
\begin{align}
\label{ESRI}
    &(\partial_t\bar{\phi})^2 \gtrsim \frac{3H^4}{8\pi^2}   \ ,
\end{align}
which almost corresponds to the limitation
\begin{align}
\label{}
    &\frac{\partial_t{\bar{\phi}}}{H}\gtrsim \frac{H}{2\pi}   
\end{align}
 for the slow roll inflation not to be in the eternal inflation phase.
Therefore, we derived the limitation for the non-eternal inflation phase from the coarse-grained Einstein equation also. This is our main result and consistent with the result using the Fokker-Planck equation in phase space \cite{Tolley:2008na}. 

We can rewrite Eq. (\ref{ESRI}) in another form. Using the expression of the power spectrum of adiabatic mode in the slow roll inflation
\begin{align}
\label{}
    &P_s(k) =\left. \frac{H^2}{8\pi^2M_{\rm pl}^2\epsilon_1}\right|_{k\simeq aH}   ,
\end{align}
we can reduce the condition (\ref{ESRI}) to
\begin{align}
\label{NE}
    &P_s\lesssim 1  \ ,
\end{align}
where we use the classical Einstein equations to evaluate the Hubble parameter and the slow roll parameter $\epsilon_1$\footnote{
We should use the classical equations since we perform all of the calculations for the cosmological perturbations on the classical background.}.
Then, we introduce the de Sitter entropy as an analog of the area law of the Bekenstein-Hawking entropy as
\begin{align}
\label{sds}
    &S_{\rm dS}:= \frac{A_{\rm dS}}{4G}= \frac{\pi}{GH^2}  \ .
\end{align}
The evolution of Hubble parameter here is described by the classical Einstein equations. 
From the de Sitter entropy bound conjecture suggested in \cite{ArkaniHamed:2007ky}, $S_{\rm dS}$ should satisfy 
\begin{align}
\label{conjecture}
S_{\rm dS}\gtrsim N\ .
\end{align} 
Variating Eq. (\ref{sds}) with respect to $N$, we obtain
\begin{align}
\label{}
    &\frac{dS_{\rm dS}}{dN} =\frac{8\pi^2\dot{\bar{\phi}}^2}{H^2} = P_s^{-1}  \ ,
\end{align} 
which almost refers to
\begin{align}
\label{}
    &S_{\rm dS}(N_{\rm end})  \sim P_s^{-1}\cdot N_{\rm end}  \ .
\end{align}
$N_{\rm end}$ is the time at the end of inflation. Combining this to Eq. (\ref{NE}), we can see that the limitation for the non-eternal inflation phase is consistent to the de Sitter entropy bound $S_{\rm dS}\gtrsim N$. 

We note also that the second Einstein equation can completely determine whether the system goes into the eternal inflation phase or not. This gives us a physical intuition. 
In the context of the effective field theory of inflation, we can understand the scalar perturbation as the Nambu-Goldstone boson associated with the spontaneous symmetry breaking of the time translation. Thus, the second order phase transition occurs in the single-field slow roll inflation. We can identify the order parameter of the phase transition to the classical background value of the field \textit{velocity} since it appears as a coefficient of the 1-particle component of the broken charge, i.e., Hamiltonian \cite{Baumann:2011ws}. 
If the fluctuation in the heat bath is so strong as to exceed the critical velocity, the NG phase will break
down and transit to highly non-equilibrium states.
The second Einstein equation in the stochastic formalism expresses this statistical description.


\subsection{de Sitter evaporation}

Next, we consider a constant potential $V_0$. In this case, we could integrate out the scalar modes of the metric perturbation. The shift symmetry of the scalar action protects the constant potential. 
If we assume that the graviton affect the system little, we could still ignore the metric perturbation.  Using Eq. (\ref{sigma}) for the constant potential, we get the coarse grained Einstein equations as
%
\begin{align}
\label{}
    &3M_{\rm pl}^2H^2 = V_0 + \frac{3H^4}{16\pi^2}  \ , \\
    &2M_{\rm pl}^2HH_{,N} =  -\frac{3H^4}{8\pi^2}   \ . 
\end{align}
Unless the slow roll inflation case, the mean value of field velocity is equal to zero and there is no phase transition. From the second equation, however, we can see that the variance resulting from the quantum fluctuation forces to collapse the classical de Sitter background.
Solving the second equation, we get
\begin{align}
\label{h}
    &H^2(N)= \left( \frac{3N}{8\pi^2M_{\rm pl}^2} + \frac{1}{H_0^2}\right)^{-1}  \ , 
\end{align}
where we set the initial values as $N_0=0$ and $H(N_0)= H_0$. 
A similar result has been obtained for the scalar field with the generic potential on exact de Sitter background \cite{Rigopoulos:2013exa}.
As time goes on, the Hubble parameter decreases in average, or, the de Sitter spacetime evaporates by the quantum effect. The quantum correction appearing in the first equation can be interpreted as the energy density of the Hawking radiation since it is proportional to the fourth power of the Gibbons-Hawking temperature $T_H^4\propto H^4$. 

We evaluated the evolution of the short wavelength part based on the classical Einstein equations (\ref{Feq}). Hence, if the classical evolution is realized at the initial time
\begin{align}
\label{h0}
    &3M_{\rm pl}^2H_0^2 = V_0   ,
\end{align} 
this solution must be held as a good approximation to describe the whole of the system.
When the classical solution significantly deviates by the quantum effect, we cannot describe the evolution of the system anymore, and the effective field theory will break down. Comparing Eq. (\ref{h}) with Eq. (\ref{h0}), we can estimate the breaking time as
\begin{align}
\label{}
    &\frac{3N_{\rm b}}{8\pi^2M_{\rm pl}^2} \sim \frac{1}{H_0^2}   \ ,  
\end{align}   
and again, this corresponds to the de Sitter entropy bound (\ref{conjecture})
\begin{align}
\label{dsn}
    &S_{\rm dS} = \frac{\pi}{GH_0^2} \sim 3N_{\rm b} \ 
\end{align}
up to $O(1)$ factor.
Therefore, the stochastic formalism can predict that the effective field theory will break down before the de Sitter spacetime completely evaporates. This is quite similar to the Page time for the black hole evaporation. The breaking time just corresponds to the de Sitter entropy bound.

According to \cite{ArkaniHamed:2007ky}, the de Sitter entropy bound implies that the Gibbons-Hawking entropy must be larger than the entanglement entropy to validate the effective field theory. In the de Sitter case, the entanglement entropy relates to the logarithm of the total Hubble patches: ${\rm ln}\,{\rm e}^{3N}=3N$. As the Hubble parameter decreases, an observer at the origin can observe more Hubble patches and the entanglement entropy increases. When the entanglement entropy reaches to the de Sitter entropy bound, we can no longer describe the system by the effective field theory on the classical de Sitter background.
After the effective field theory breaks down, the system would enter in a highly non-equilibrium state. In such a regime, we have to choose another vacuum besides the de Sitter background or consider a full evolution of gravity sector.

We pointed out that the de Sitter entropy bound conjecture for the effective field theory can apply not only for the slow roll inflation but also for the scalar field on de Sitter background, using the stochastic formalism and the coarse-grained Einstein equations. 
Then, we discuss discrepancies between our approach and the results based on the quantum field theory. Usually, the equation of state parameter of radiation is $1/3$, while that of the scalar field on the de Sitter background is $1$.
Thus, in the stochastic formalism, we treat the (would-be) radiation in a different way from the thermal field theory as in \cite{Markkanen:2016jhg, Markkanen:2017abw}. The breaking time of de Sitter background also differs from the results in \cite{Markkanen:2016jhg, Markkanen:2017abw, Matsui:2019tah, Aalsma:2019rpt} by $O(10^2)$ numerical factor. 
We expect that the stochastic formalism is the first order approximation of the quantum field theory so that they would correspond in some limits as \cite{Rigopoulos:2016oko}. Thus, we need to search why the differences between the two approaches appear for future work.

\section{Summary}\label{5}


We have introduced the stochastic formalism with the three assumptions for the canonical scalar field and the general relativity.
We derived the Fokker-Planck equation leaving the second order time derivative of the long wavelength part.
We obtained the equilibrium distribution of the field velocity and reconfirmed that the second fluctuation-dissipation theorem is held.
Then, by using the coarse-grained Einstein equations, we evaluated the quantum effect from the short wavelength part to the long wavelength part.
For the slow roll potential case, we can derive the condition for the non-eternal inflation phase from the second Einstein equation (\ref{2nd}). The second Einstein equation describes that if the fluctuation consisting of the short wavelength part is too strong, we cannot maintain the Nambu-Goldstone phase generated by the long wavelength part. The borderline of the phase transition almost corresponds to the de Sitter entropy bound.   
For the constant potential case, on the other hand, the fluctuation causes the Hawking radiation-like behavior for the long wavelength part, and the de Sitter spacetime evaporates. However, 
as the evaporation proceeds and the horizon size becomes larger, the effective field theory based on the classical de Sitter background will break down. The breaking time corresponds to the de Sitter entropy bound again.
We have found that for both of the slow roll potential case and the constant potential case, the constraints to the theories derived from the coarse-grained Einstein equations 
correspond to the de Sitter entropy bound well.


Finally, we comment on quantum correction to observables in quasi-de Sitter spacetime.
We expect naively that as for the quantum correction, we need the renormalization of the short wavelength part only since the long wavelength part would classicalize and lose its quantum property. If we regard the results from the stochastic formalism as the renormalization to the vacuum expectation value of the scalar field, from the second Einstein equation (\ref{2nd}), we may estimate that the quantum correction is proportional to $(H/M_{\rm pl})^2$. If the energy scale of the Hubble parameter is much below than the Planck scale, the correction is well suppressed and it would be sufficient to use the usual classical results to compare with observational values, especially for the inflation. We should investigate the behavior of the quantum correction from the full quantum field theory and compare the results with those of the stochastic formalism also.   

\acknowledgments
The author's work is partially supported by Leung Center for Cosmology and Particle Astrophysics, National Taiwan Univ. (FI121).

\appendix
\section{Correlation of noise}\label{A}

We calculate the correlation of the noise $\Delta \hat{W}$. The noise consists of the short wavelength part only so that we should take a quantum average. We derive the correlation at the same points since we need the one-domain probability distribution function to compute the average value of energy-momentum tensor. Up to the leading order of slow roll parameters, we obtain
\begin{align}
\label{cor}
    \langle \Delta \hat{W}^2\rangle&= \int^{N+\Delta N}_NdN_1dN_2
                     \ \langle0| \hat{\xi}_S(N_1)\hat{\xi}_S(N_2)|0\rangle    \\
      &=\int^{N+\Delta N}_NdN_1dN_2 \left(\frac{H}{2\pi}\right)^2
         (\partial_{N_1}+3)(\partial_{N_2}+3)\delta(N_1-N_2) \\
      &\simeq  \left(\frac{H}{2\pi}\right)^2\left\{ 9\Delta N + \int^{N+\Delta N}_NdN_1
      dN_2 \left[6\partial_{N_1}\delta(N_1-N_2) +
    \partial_{N_1}\partial_{N_2}\delta(N_1-N_2)\right]\right\} \ ,
\end{align}
where 
\begin{align}
\label{}
    \int^{N+\Delta N}_NdN_1dN_2 := \int^{N+\Delta N}_NdN_1\int^{N+\Delta N}_NdN_2 \ .
 \end{align}
To handle the derivatives of Dirac's delta function, we define it with a regularization as
\begin{align}
\label{}
    \delta (z) &:= \lim_{\epsilon \to 0}\int \frac{d\omega}{2\pi}{\rm e}^{i\omega z - \epsilon\omega^2}   \\
    &= \lim_{\epsilon \to 0}\int \frac{d\omega}{2\pi}{\rm e}^{-\epsilon\omega^2}
                \sum_{n=0}^\infty\frac{i^n}{n!}(\omega z)^n \ ,
\end{align}
where the integration covers all of $\omega$-space.
By using this definition, we obtain
\begin{align}
\label{}
    &\int^{N+\Delta N}_NdN_1dN_2 \partial_{N_1}\delta(N_1-N_2) \nonumber \\
    =&\lim_{\epsilon  \to 0}\int^{N+\Delta N}_NdN_1dN_2\int\frac{d\omega}{2\pi}
       i\omega\, {\rm e}^{i\omega(N_1-N_2)-\epsilon \omega^2} \nonumber \\
    =&\lim_{\epsilon \to 0}\int\frac{d\omega}{2\pi}\frac{1}{i\omega}\,
          [{\rm e}^{i\omega\Delta N}+{\rm e}^{-i\omega\Delta N}-2]{\rm e}^{-\epsilon\omega^2} 
           \nonumber \\
    =&\lim_{\epsilon \to 0}[{\rm e}^{i0\cdot\Delta N}+{\rm e}^{- i0\cdot\Delta N}-2]{\rm e}^{-\epsilon\cdot0^2} = 0 \ ,
\end{align}
and
\begin{align}
\label{div}
    &\int^{N+\Delta N}_NdN_1dN_2 \partial_{N_1}\partial_{N_2}\delta(N_1-N_2) \nonumber \\
    =&\lim_{\epsilon  \to 0}\int^{N+\Delta N}_NdN_1dN_2\int\frac{d\omega}{2\pi}
       \omega^2\, {\rm e}^{i\omega(N_1-N_2)-\epsilon \omega^2} \nonumber \\
    =&\lim_{\epsilon \to 0}\int \frac{d\omega}{2\pi}
        \left[ 2 -{\rm e}^{i\omega\Delta N}-{\rm e}^{-i\omega\Delta N} \right]{\rm e}^{- \epsilon\omega^2} \nonumber \\
    =&\lim_{\epsilon \to 0}2\int \frac{d\omega}{2\pi} \sum_{n=1}^\infty\frac{(-1)^n}{(2n)!}( \Delta N)^{2n}\omega^{2n}{\rm e}^{- \epsilon\omega^2}  \nonumber \\
         =&\lim_{\epsilon \to 0}\sum_{n=1}^\infty\frac{\sqrt{\pi}(-1)^{n}(2n-1)!!}{2^n(2n)!}
              \frac{\Delta N^{2n}}{\epsilon^{n+\frac{1}{2}}} \ .
\end{align}
The second one diverges if $\Delta N$ is finite. This divergence might be caused by the quantum average for the composite operator at the same time.
Here, instead of going to the details, we deal with the divergence just by controlling how $\epsilon$ approaches to zero. To suppress the contribution from Eq. (\ref{div}) to the correlation function (\ref{cor}), we need
\begin{align}
\label{}
    &\lim_{\epsilon,\,\Delta N\to 0}\frac{\Delta N^{2n-1}}{\epsilon^{n+\frac{1}{2}}} \to 0    \quad {\rm for}\ {}^\forall n\ .
\end{align}
Assuming $\epsilon\propto (\Delta N)^\gamma$ where $\gamma$ is a constant parameter, $\gamma$ must be satisfied with
\begin{align}
\label{}
    &\gamma <\frac{2n-1}{n+\frac{1}{2}} \quad {\rm for}\ {}^\forall n   
\end{align}
which implies $\gamma < \frac{2}{3}$.
Therefore, to remove the divergence, we take $\epsilon$ as it approaches zero more slowly than $\Delta N^{2/3}$. This regularization scheme seems a bit artificial. By using this scheme, however, we can obtain the white noise only, which corresponds to the usual stochastic formalism. Thus, we adopt this scheme in this article. We finally obtain the regularized correlation function as 
\begin{align}
\label{}
    &\langle \Delta W^2\rangle \simeq \left(\frac{3H}{2\pi}\right)^2\Delta N 
\end{align}
up to $O(\Delta N)$.

\bibliographystyle{apsrmp}

\end{document}